\documentclass[sigconf]{acmart}
\usepackage{meta/commands}

\AtBeginDocument{%
  \providecommand\BibTeX{{%
    \normalfont B\kern-0.5em{\scshape i\kern-0.25em b}\kern-0.8em\TeX}}}

\copyrightyear{2024}
\acmYear{2024}
\setcopyright{acmlicensed}\acmConference[CHI '24]{Proceedings of the CHI Conference on Human Factors in Computing Systems}{May 11--16, 2024}{Honolulu, HI, USA}
\acmBooktitle{Proceedings of the CHI Conference on Human Factors in Computing Systems (CHI '24), May 11--16, 2024, Honolulu, HI, USA}
\acmDOI{10.1145/3613904.3642853}
\acmISBN{979-8-4007-0330-0/24/05}

\begin{document}

\title[DiaryHelper]{DiaryHelper: Exploring the Use of an Automatic Contextual Information Recording Agent for Elicitation Diary Study}

\author{Junze Li}
\email{junze.li@connect.ust.hk}
\affiliation{%
 \institution{The Hong Kong University of Science and Technology}
 \city{Hong Kong}
 \country{China}
}

\author{Changyang He}
\email{cheai@connect.ust.hk}
\affiliation{%
 \institution{The Hong Kong University of Science and Technology}
 \city{Hong Kong}
 \country{China}
}

\author{Jiaxiong Hu}
\email{hujx@ust.hk}
\affiliation{%
 \institution{The Hong Kong University of Science and Technology}
 \city{Hong Kong}
 \country{China}
}

\author{Boyang Jia}
\email{boyangj@alumni.cmu.edu}
\affiliation{%
 \institution{Tsinghua University}
 \city{Beijing}
 \country{China}
}

\author{Alon Halevy}
\email{halevy@amazon.com}
\affiliation{%
 \institution{Amazon Web Services}
 \city{Palo Alto}
 \country{USA}
}
\authornote{This work was done while the author was at Meta AI.}

\author{Xiaojuan Ma}
\email{mxj@cse.ust.hk}
\affiliation{%
 \institution{The Hong Kong University of Science and Technology}
 \city{Hong Kong}
 \country{China}
}
\authornote{Corresponding author}

\renewcommand{\shortauthors}{Junze Li et al.}

\begin{abstract}
Elicitation diary studies, a type of qualitative, longitudinal research method, involve participants to self-report aspects of events of interest at their occurrences as memory cues for providing details and insights during post-study interviews. However, due to time constraints and lack of motivation, participants’ diary entries may be vague or incomplete, impairing their later recall. To address this challenge, we designed an automatic contextual information recording agent, \tool\!, based on the theory of episodic memory. \tool can predict five dimensions of contextual information and confirm with participants. We evaluated the use of \tool in both the recording period and the elicitation interview through a within-subject study (N=12) over a period of two weeks. Our results demonstrated that \tool can assist participants in capturing abundant and accurate contextual information without significant burden, leading to a more detailed recall of recorded events and providing greater insights.
\end{abstract}





\maketitle
\section{Introduction}

A diary study is a contextual, longitudinal, qualitative research method that relies on participants' self-recorded data to inspect their behaviors, activities, and experiences over time \cite{Rieman1993thediarystudy, bolger2003diary}. 
Diary studies are widely used in domains such as health, psychology, and education because of the rich and detailed data collection they enable \cite{strahler2020associations,dalton2018independent,sonnentag2001work,conner2015creative,PETERSON201582,bakker_student_2015,dantec2008homeless}. For Human-Computer Interaction (HCI) research, diary studies provide an in-depth understanding of users' real-world interaction with technology, providing valuable insights for design improvements in users' natural environments~\cite{czerwinski2004taskswitching, van2006interviewviz}.
Diary studies can be divided into two types, namely feedback studies and elicitation studies \cite{Carter2005media}, based on the data recording methods.
In feedback studies, participants provide information about all required aspects of an event related to the study topic immediately by answering a set of pre-defined questions \cite{czerwinski2004taskswitching,hess2009richmedia,Tuomas2010location}.
Since participants' responses are structured and collected promptly and fully after they perceive the events, the recorded diary entries are not affected by memory decay \cite{Timothy2016memorydecay}. 
However, replying to a questionnaire may distract participants from their ongoing tasks, which tends to cause a high dropout rate in these diary studies \cite{Carter2005media, Chong2015cuenow}. 
Elicitation studies, on the other hand, ask participants to only capture some aspects of the events of interest when they happen. The captured information is supplemented by an elicitation interview that happens after the recording period, and includes going through their gathered data, recalling the events, and providing more details and insights relevant to the research topic.
Logging is more unobtrusive and flexible in this type of diary study, as participants can do it at their discretion, and there is usually no restriction on the data aspects and formats to record in situ \cite{van2006interviewviz,sun2011user,gabridge2008information}.
The downside of elicitation studies is that participants may forget some details of the reported events by the time of the elicitation interview, especially when their recordings are vague or incomplete \cite{Carter2005media, Swallow2022mobilitybarriers, Gorm2017photo}.

In general, an elicitation study is more suited for research topics in which the number of events to log can be large because it is less burdensome for participants than a feedback study \cite{petersen2002usability,Carter2005media}. People can simply log snippets of information about each event and use them as memory cues in the later interview. They are able to record data rather conveniently with the help of diverse media available (\eg a few words, a photo, a voice message, a vlog \etc) \cite{Carter2005media,Suh2018videochat,Brandt2007snippet,palen2002voicemail,metz2021vlog,sun2011user}.
The challenge lies in how to enrich such memory cues with contextual information to foster recall.
Prior studies tried to encourage participants to add some annotations, such as thoughts or short descriptions of recorded events, for each dairy entry to jog their episodic memory in the elicitation interview \cite{Carter2005media, Gorm2017photo, Suh2018videochat}. 
However, they found that their participants tended to leave very few, even zero, annotations \cite{Carter2005media, Suh2018videochat}. 
To address this issue, some researchers developed lifelogging systems or deployed other devices to automatically collect in situ information, such as GPS locations or objects photographed for augmenting participants' self-reports, which was shown to help them memorize more details of past events \cite{Gouveia2013footprint, Kalnikaite2010lifeloglocation, hodges2011sensecam}. 
However, this approach requires additional efforts of taking or wearing extra devices and thus may not be feasible for diary studies lasting for a long period of time. 
It also limits the scalability of the study due to the cost of devices.


In this paper, we introduce \tool\!, a tool that uses generative AI techniques to enable capturing more details of events with less burden on participants, resulting in less retrospection bias in the elicitation interview. \tool can be integrated into the diary logging platform and predicts for each recorded event five dimensions of contextual information that are known to be helpful for episodic memory \cite{tulving_1984,Lee2007emi}: time, location, emotion, people, and activity.
\tool leverages cloud services to understand the textual and/or visual contents in a diary post and in particular, uses a large language model (LLM) to predict the possible labels of the five dimensions. To mitigate the hallucination problem that is common to LLMs, we pre-specify the space of possible predicted labels. 
Participants can review the memos generated by \tool at their convenience and modify the information when necessary. 

To evaluate \tool\!, this work aims to explore the following two research questions:

\textbf{RQ1} In the recording period, can \tool effectively assist participants in capturing abundant and accurate contextual information for their recorded diary entries?
 
\textbf{RQ2} In the elicitation interview, can study participants recall more details for each recorded event and provide more insights based on the memo generated by \tool\!?

To answer these questions, we recruited 12 people to join a two-week diary study.
We used Slack as the diary logging platform for the study, which is accessible on both laptops and smartphones. 
Slack supports four recording modalities, text, audio, image, and video, thereby enabling participants to capture events in different situations.  
\tool was embedded into Slack as a chatbot and integrated into every private Slack channel created for each participant to post data. 
The study was conducted in a within-subject manner, in which participants recorded diary entries only by themselves for one week and with the assistance of \tool for another week. 
We counterbalanced the order of the two logging conditions. 
Through the user study, we found that users generally considered five dimensions of contextual information generated by \tool to be accurate and consistent with in situ information. We also found that \tool decreased the burden of diary recording, which promoted participants' recording willingness.
During the elicitation interview, users with the assistance of \tool\! demonstrated improved recall ability regarding the recorded events, such as elevated recall levels of emotions and activities. They provided more detailed retrospective descriptions of recorded events, which also brought more insights and discussions regarding the study topic.



The main contributions of this paper are threefold:  
\begin{itemize}
    \item \tool\!, an agent to assist participants in recording contextual information automatically for elicitation diary studies.
    \item A study demonstrating the effectiveness of \tool in improving participants' willingness to record with less burden and promoting recall and discussion in elicitation interviews.
    \item Design implications on how to customize \tool to a specific diary study topic, and more generally, how to use generative AI techniques to help us capture more of the subjective experiences in our lives.  
\end{itemize}
\section{Literature Review}
In this section, we summarized literature adopting diary study methods, the challenges and difficulties emerged in conducting diary studies, and improved diary study methods. Since the recall process is important in elicitation diary studies, we also surveyed techniques created for facilitating recall.

\subsection{Diary Study Implementations}

Brown \etal implemented diary study methods to explore how people capture information with the emergence of new devices \cite{brown2000information}.
Carter \etal reported a diary study to discover how people make transit decisions in their daily lives, where recruited participants made a phone call when they made a transit decision \cite{Carter2005media}.
Loeffler conducted a photo elicitation diary study to explore the meanings and benefits of participating in outdoor activities \cite{loeffler2004outdoor}.
Dantec \etal investigated how technologies affect homeless population through a photo diary study, and demonstrated the potential of diary study to collect rich behavioral data \cite{dantec2008homeless}.
Hong \etal implemented diary probes to document illness experience of adolescent patients and their parental caregivers \cite{hong2020adolescent}.
Overdevest \etal adopted diary study methods to explore human's behavior of embodied remembering \cite{Overdevest2023embodied}. 
Generally speaking, diary studies are widely used to observe and investigate human behaviors, and collect abundant longitudinal data of particular user groups.
\rr{Previous studies also reported participants' recording habits. Steves \etal found that most of the participants recorded authentication related events when they were not working \cite{steves2014report}. Brandt \etal discovered participants' unwillingness to record while they were mobile or active \cite{Brandt2007snippet}.
Meanwhile, diary recordings can also affect participants' behaviors and insights.
Isaacs \etal revealed that the technology mediated recordings and reflections facilitated participants' well-being, such as happiness and satisfaction \cite{Isaacs2013echo}.
Gorm \etal mentioned that the photos taken during the logging period sparked participants' insights and reflections in the interview, and they retrospected their past experiences when looking at the photos \cite{Gorm2017photo}.}

\subsection{Challenges and Difficulties in Diary Study}
Carter \etal discovered that diverse choice for data capture could lower participants' recording burden, but some supplementary contextual information were useful for participants' recall and discussion \cite{Carter2005media}.
A research guide proposed by Singh \etal indicated some unique challenges in diary studies, such as duration, information collection, and keeping users interested \etc \cite{anjeli2013guide}.
Parnell \etal conducted a diary study to discover distracted driving behaviors in vehicles, and they found that some interactions were hard to record and participants might have recollection bias based on the recorded data \cite{PARNELL20201}.
Gorm \etal reported some recall and memory issues in the elicitation interview, and suggested that participants could leave some text along with the photos when reporting the captured data \cite{Gorm2017photo}.  
During the photo diary study to collect mobility barriers faced by aged adults, Swallow \etal mentioned that participants might forget what they had taken and the reasons for taking such photos \cite{Swallow2022mobilitybarriers}. 
It is a trade-off between the convenience of recording and the richness of recorded data, so our work aims at balancing these two aspects to assist participants in both the recording period and elicitation interview.

\subsection{Improved Diary Study Methods}
\rr{To lower the participants' recording burden in situ, researchers proposed methods that involved delayed completion of diary entries and quick data capture.} Brandt \etal let the participants capture some snippets (bits of text, audio, or photos) and upload them to a server through mobile phones immediately, then they could access a website to construct detailed and structured diary entries at a convenient time \cite{Brandt2007snippet}.
Chong \etal proposed the idea of delayed reflection for diary studies, in which participants could squeeze a sensor to capture some cues quickly and reflect on these cues later \cite{Chong2015cuenow}.
Zhang \etal examined the usefulness of unlock journaling to support lightweight in situ self-report \cite{zhang2016unlock}.
\rr{However, these methods may lead to the loss of detailed contextual information related to the recorded events.}
\rr{Another type of improved methods focused on the elicitation interview after recording period.}
House designed a visualization method to organize photos in different dimensions to trigger more detailed interviews \cite{van2006interviewviz}. 
Feng \etal designed an enhanced photo elicitation diary method to aid participants' recall and obtain detailed findings \cite{feng2019enhanced}.
\rr{These methods can facilitate  participants' discussion to provide more insights in the interview, but the problems and difficulties of diary recording are not mitigated.}
Instead, our work focuses on the design of recording rich contextual information automatically based on the understanding of recorded data empowered by LLM.

\subsection{Techniques for Facilitating Recall}
\rr{Episodic memory \cite{tulving1993episodic} involves storing and retrieving information about people's daily experiences. Previous research about episodic memory provided a theoretical basis to design systems and methods for facilitating recall \cite{Sellen2007memorysensecam,Lee2007emi,Kalnikaite2010lifeloglocation,Gouveia2013footprint,Le2016videosum}.}
For example, Sellen \etal used a wearable camera, SenseCam, to capture daily data and investigated its effect of supporting people's memory \cite{Sellen2007memorysensecam}.
Lee \etal investigated how to extract efficient memory cues when using lifelogging techniques for people with episodic memory impairment \cite{Lee2007emi}.
Kalnikaitė \etal focused on the effect of location information on people's memory in lifelogging systems \cite{Kalnikaite2010lifeloglocation}.
Gouveia \etal designed a lifelogging system, Footprint Tracker, to review four different contextual information as memory cues to facilitate participants' recall and reflection of daily activities \cite{Gouveia2013footprint}.
Due to the large amount of data captured by lifelogging systems, Le \etal studies the impact of video summarization for recall and how to summarize videos for memory augmentation \cite{Le2016videosum}.
These techniques are mainly based on the lifelogging devices, which increase the intrusiveness of study and the difficulty of processing log data.
\rr{In addition to the lifelogging systems, some other systems were also constructed to facilitate people's memory retrieval.}
Sas \etal designed AffectCam system to help participants capture high arousal photos for triggering richer recall \cite{sas2013affectcam}.
Isaacs \etal showed that the Echo system can support not only recording the emotion of events but also a reflection on past events \cite{Isaacs2013echo}.
In our design, we adopted the effective memory cues reported in these works as the contextual information to record, and designed a lightweight approach to diary recording.
\section{System Design}
In this section, we introduce the selection of contextual information to record based on the episodic memory theory, the implementation details of \tool\!, and the user interface of \tool\!.

\subsection{Selection of Contextual Information}\label{sec:5dim}
Episodic memory \cite{tulving1993episodic} is a type of long-term memory that includes the recollection of specific events and personal status, such as time, location, associated emotions, and people involved.
It is vulnerable to memory decay, and can be affected by various factors such as age, pressure, and neurological status \cite{Timothy2016memorydecay}.
In the elicitation interview of a diary study, participants need to reconstruct their episodic memories based on their recorded data.
As Tulving mentioned, the cues related to ``\textit{who, what, where, and when}'' play a pivotal role for episodic memory recall \cite{tulving_1984}. 

Following the episodic memory theories and existing work about episodic memory enhancement \cite{Gouveia2013footprint,Le2016videosum,Lee2007emi,hodges2011sensecam,Daniel2004drm}, we selected five dimensions of contextual information to record in \tool\!, which are time, location, emotion, people, and activity.

\subsection{Implementation}\label{sec:implementation}
\tool was embedded into the Slack platform\footnote{\url{https://slack.com/}}, a widely recognized workplace messaging application, where each participant could post recorded diaries in a private channel. 
\rr{Slack is also available on different devices (\eg cell phones, desktops, and tablets), and supports flexible integration of toolkits and apps.}
Then the back-end program processed the diary data and predicted the five dimensions of contextual information.  
This implementation enables streamlined diary logging while preserving abundant contextual information by: conversational interactions through a familiar user interface, seamless logging with minimal waiting times, and automated annotations with a diverse range of predictions.

\subsubsection{Front-End Integration with Slack}
Participants can interact with \tool by posting recorded diaries within their private diary channel, ensuring privacy and confidentiality.
\tool supports four modalities of recording: text, audio, image and video.
When participants submit a diary entry, \tool responds with an acknowledgment message and saves the submitted data into the database.
Then a memo form including contextual information predicted by \tool is generated for participants' checking.
After confirming the content of the memo, \tool saves the confirmed memo form in JSON format.
All the interactions on the front-end interface were deployed with Slack API, which is an efficient method to build integrated applications in Slack.  

\subsubsection{Multimodal Data Processing}
\label{multimodaldata}
Participants' posts are forwarded to the back-end Python-based program in real-time.
We leveraged cloud service APIs to process and understand the content of recorded data. 
Except the text-based data, we implemented the following APIs to process images, videos, and audios:

\begin{itemize}
    \item Image Processing: We employed the Microsoft Azure Computer Vision service to transform images into textual descriptions encompassing categories, brands, descriptions, objects, and tags detected in the image.
    \item Video Processing: Google Cloud Video Intelligence service was utilized for video content analysis, capturing items and events portrayed in the video clips.
    \item Audio Processing: Audio clips were converted into text messages using Google Cloud Speech service.
\end{itemize}

The criteria of API selection is based on performance and response time assessments, which are crucial considerations to ensure the accuracy of the contextual information and limit the participants' waiting time.

\subsubsection{Contextual Information Prediction by LLM}
We adopted GPT-3.5 to predict five dimensions of contextual information. 
\rr{The temperature value was set to 0.7 to balance the reliability and creativity of generated contents \cite{liu2023creative}.}
The template of message sent to GPT-3.5 consists of three parts: 1) the basic introduction of the task, 2) the features extracted by cloud service APIs, 3) the detailed instruction of prediction.

\begin{itemize}
    \item[\textbf{1)}] the basic introduction of the task. It indicates the role of GPT-3.5 as an experienced diary study researcher, and the purpose of recording contextual information.
    \item[\textbf{2)}] the content of the recorded diary data. The textual contents (\ie text message and transcription of audio clip) are inserted in the template directly. For visual content (\ie image and video), we inserted the object tags and descriptions generated by the APIs introduced in Section \ref{multimodaldata} into the template.
    \item[\textbf{3)}] the instructions for contextual information prediction. The time information was extract from the metadata of diary message. To mitigate the hallucination issues of LLMs \cite{ji2023hallucination}, we pre-specified the space of possible predicted labels. For the location information, we instructed GPT-3.5 to predict from the point of interest location categories in Google Maps. The emotion tag was selected from ``\textit{positive}'', ``\textit{neutral}'' and ``\textit{negative}''. The considered categories of people were ``\textit{alone}'', ``\textit{families}'', ``\textit{friends}'', ``\textit{colleagues}'' and ``\textit{acquaintances}''. For the activity descriptions, we limited the length no more than 151 characters to meet the display requirements in Slack. 
\end{itemize}

To improve the stability of the GPT-3.5's output, we also explained the output format explicitly and provided some concrete examples in the prompt template, which was demonstrated useful in previous work \cite{ramesh2021zeroshot, brown2020fewshot}.
We present an exact prompt for image-based diaries in Appendix \ref{app:prompt}.


\subsection{User Interface}
The user interface of \tool is shown in Figure \ref{fig:interface}.
\tool was installed in each private channel to collect participants' recorded diaries and assist them in capturing contextual information.
The process of recording one diary entry with \tool includes following five steps:

\begin{figure*}[!ht]
    \includegraphics[width=15cm]{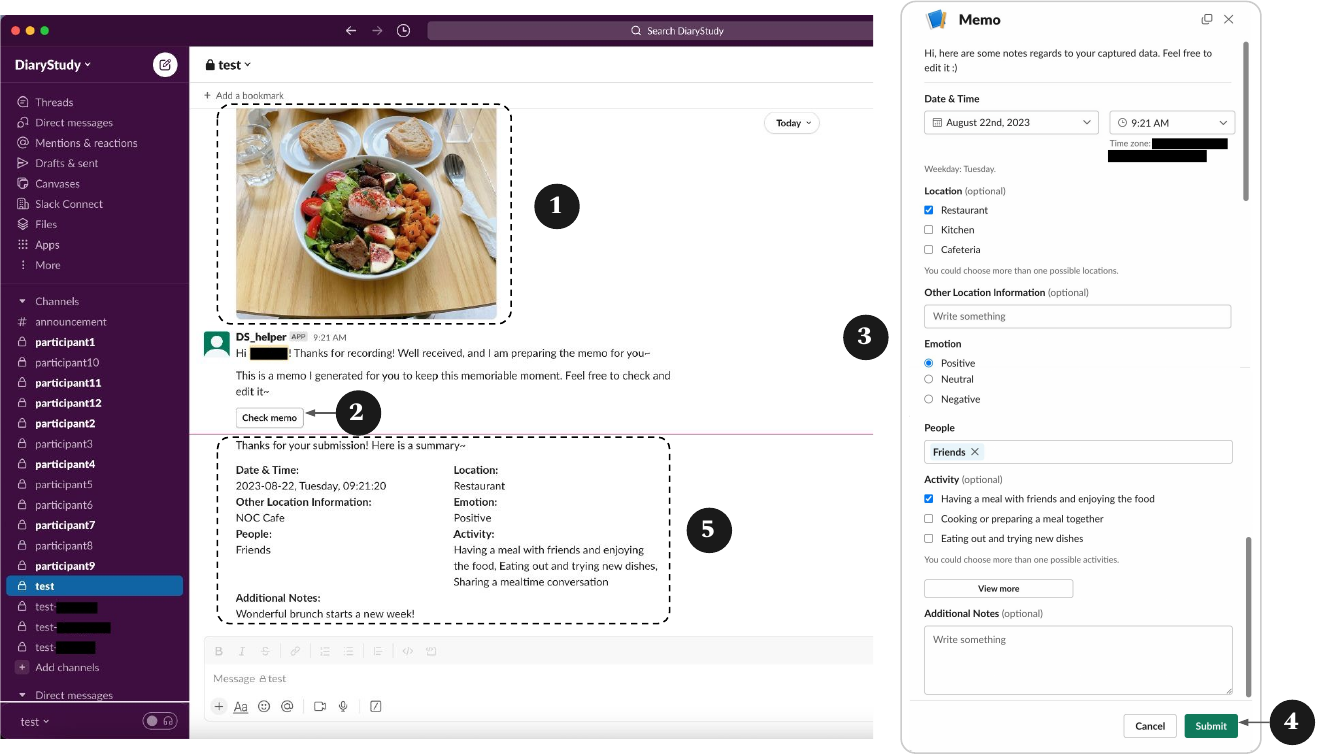}
    \caption{The interface of \tool\!.}
    \label{fig:interface}
    \Description{Figure 1 illustrates a Slack Channel interface with \tool installed. Participants can initiate diary recording via sending a message in the input box. In this example, the user posted an image of a brunch. \tool acknowledged the message was well received by sending a reply. It then posted a memo form in the channel for the user to check. After verifying the information in a pop-up box (shown on the right hand side), \tool sent out a confirmation message in the chat, which included all the confirmed contextual messages related to this diary entry.}
    \centering
\end{figure*}

\begin{description}
    \item[\textcircled{1} Participants' Diary Input:] Participants can initiate the diary recording by posting a multimodal message via the bottom input box. They could record events of interest at once or upload some data captured previously.

    \item[\textcircled{2} Immediate Acknowledgment:] After receiving a message, \tool promptly confirms the message was received and acknowledges participant's diary recording. \tool then sends a message indicating a memo is generated. Participants could click the ``\textit{Check Memo}'' button to annotate contextual information at their convenience.

    \item[\textcircled{3} Memo Form:] The memo form appears as a pop-out window in Slack, providing participants with predicted contextual information. The memo form is designed to reduce the burden of filling out for participants and includes the following components:
\begin{itemize}
    \item Date \& Time: Automatically selected based on the metadata of the message. Participants can adjust it if necessary.
    \item Location: Three location tags predicted by \tool, listed in order of probability. The first option is selected by default, but participants can modify and select other options. An optional text input box is provided for participants to leave additional location information if necessary.
    \item Emotion: The predicted emotion tag by \tool is selected by default, and participants can change it. These options are exclusive, preventing multiple selections.
    \item People: The predicted category of people involved is selected by default. Participants can delete this option or add more categories as needed.
    \item Activity: Similar to location and emotion information, the first option with highest probability is pre-selected. Three options are provided initially and participants can click ``\textit{View More}'' button to explore more potential activity descriptions. A text input box below allows participants to leave some additional contextual information if necessary.
\end{itemize}

    \item[\textcircled{4} Review and Confirmation:] After checking and editing the memo form, participants can click the ``\textit{Submit}'' button to keep this memo as the contextual information recorded for this diary entry.

    \item[\textcircled{5} Summary:] Finally, \tool confirms the receipt of the memo and presents the recorded contextual information in a summary message, which is clear and structured for participants to review in the elicitation interview. 
\end{description}

The memo generated by \tool provides a flexible and lightweight way of recording five dimensions of contextual information for each diary entry. 
Participants can also append any additional contextual information in the thread of each message at any time.

\section{Evaluation}
To explore the effectiveness of \tool in diary studies, we conducted a two-week elicitation diary study in a within-subject manner. 
In this section, we introduce the details of study procedures and corresponding measurements.

\subsection{Study Procedures}\label{sec:exp_pip}
With the approval of our institution’s IRB, we recruited 12 participants (3 females and 9 males; age range 22-28, M=25.58, S.D.=2.14; employment status six student, six employee; working mode two remote, two hybrid, eight onsite; summarized in Table \ref{tab:demographic}) into our diary study through online advertisement and word-of-mouth. 
None of the participants suffer from memory impairment issues and they can communicate in English proficiently. 
\rr{Most participants (8/12) have used Slack before, and all participants have used similar applications, e.g. Microsoft Teams, Discord, etc. No participant reported difficulties when using Slack in the diary recording period.}

\begin{table}[!ht]
    \centering
    \caption{Demographics of all the participants, including participants’ ID, gender, age, their employment status and working mode.}
    \begin{tabular}{ccccc}
    \toprule
    ID & Gender & Age & Employment Status & Working Mode  \\
    \midrule
    1 & Male & 24 & Student & Remote\\
    2 & Male & 27 & Employee & Onsite\\
    3 & Female & 28 & Employee & Onsite \\
    4 & Male & 27 & Employee & Onsite \\
    5 & Male & 28 & Employee & Onsite \\
    6 & Male & 27 & Student & Onsite \\
    7 & Male & 22 & Student & Remote \\
    8 & Female & 22 & Student & Hybrid \\
    9 & Male & 27 & Student & Hybrid \\
    10 & Male & 26 & Employee & Onsite \\
    11 & Male & 23 & Student & Onsite \\
    12 & Female & 26 & Employee & Onsite \\
    \bottomrule
    \end{tabular}
    \label{tab:demographic}
\end{table}

\begin{figure*}[!ht]
    \includegraphics[width=15cm]{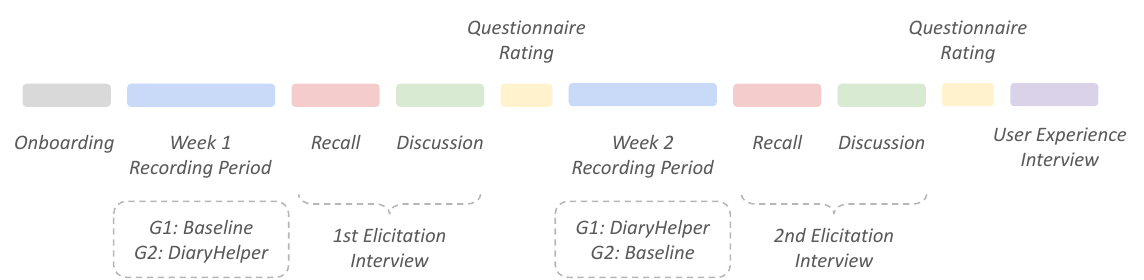}
    \caption{The general pipeline of our study (participant group 1/2 are denoted as G1/2).}
    \label{fig:timeline}
    \Description{Figure 2 presents a chronological pipeline of our study, consisting of ten stages. Participants will start with an on-boarding event, followed by the first recording period, elicitation interview including a recall session and a discussion session, and a questionnaire rating in week one. The same procedure is repeated in week two, starting with the second recording period, followed by another elicitation interview, and questionnaire rating. Finally, the study ends with a user experience interview.}
    \centering
\end{figure*}

\rr{Our study pipeline is shown in Figure \ref{fig:timeline}. During the onboarding of the study, we introduced our study purpose and procedures, and provided a detailed tutorial about using Slack and \tool to record diary entries.
They were also suggested to annotate some contextual information to support their later recall in the elicitation interview. We recommended the five dimensions of contextual information mentioned in Section \ref{sec:5dim} for them to record.}
The topic of diary study we selected is ``\textit{How do individuals manage their work-life balance in flexible working mode}''. 
This topic is multifaceted and therefore participants can easily find events of interest in their daily lives.
However, it requires nuanced and detailed dairy recordings of seemingly ordinary daily events. 
Otherwise, participants may struggle to reflect on the study topic and provide more insights during the elicitation interview.
Therefore, the topic is well suited for testing the performance of \tool\!.
\rr{In each elicitation interview, participants needed to recall each recorded event in the past week in  chronological order by looking at the diary entries, discuss the relation between the recorded diaries and study topic, and reflect on the study topic to provide more insights.
After experiencing the system in each week, participants rated a questionnaire at the end of the elicitation interview (to be detailed in Section \ref{sec:exp_eval}).
In the final user experience interview, we asked open-ended questions about participants' recording habits we observed and the difference in user experience between these two systems, such as recording burden, recording willingness, and assistance for recall.
All the interview sessions were conducted remotely and video recorded. The video recordings were transcribed for further analysis.}

We considered the original Slack platform without \tool installed as the Baseline system. 
Half of the participants (G1: P1-P6) recorded diaries using the Baseline system for one week followed by the first elicitation interview, then recorded diaries using \tool for another week followed by the second elicitation interview.
To counterbalance the experiment, the other half of the participants (G2: P7-P12) followed the same procedure, but in the opposite order (using \tool in the first week). 
\rr{The two groups of participants were counterbalanced randomly and there was no stratification. Both groups included participants with different gender, employment status and working mode.}






    
    

\subsection{Measurements}\label{sec:exp_eval}
We mainly assessed the systems in three dimensions: system usefulness for diary recording, system usefulness for elicitation interview, and system usability.
The design of questionnaire and semi-structured user experience interview were based on these assessments.

\textbf{System Usefulness for Diary Recording.} Apart from the statistical results of recorded diary data, we also asked participants about three aspects of the questionnaire: 1) whether they would like to record diaries frequently (\textit{frequency}); 2) whether they would like to record diaries immediately when some events of interest happened (\textit{timeliness}); 3) whether they would like to record diaries in different modalities (\textit{modality}).  

\textbf{System Usefulness for Elicitation Interview.} The purpose of diary studies is not only collecting longitudinal behavior data, but also providing insights on the study topic.
Therefore, we evaluate the participants' performance during the interview in three levels: 1) the recollection of recorded diaries (\textit{recall}); 2) the description of the relation between recorded diaries and study topic (\textit{relation}); 3) the reflection on the study topic (\textit{reflection}).

Referring to the existing work about measuring the quality of recall \cite{sas2013affectcam,Le2016videosum}, we determined a criteria to score the participants' recall in five dimensions: time, location, people, emotion and activity.

We transcribed all the interview audio files and filtered out the contents regarding the recollections of past events.
Then we randomly sampled 50 recollections. 
After initial discussion about the criteria, two authors scored these sampled recollections independently in the scale of 0, 1 and 2. 
The agreement ratio of these two authors' score in all five dimensions were higher than 0.9, which indicated a high consistency.
After discussing and fixing all the disagreements, the criteria to evaluate recall was finalized:

\begin{itemize}
    \item Time: 0 (not mention or unable to remember), 1 (mention some incomplete descriptions, such as one day, morning, or afternoon), 2 (remember a clear time point).
    \item Location: 0 (not mention or unable to remember), 1 (mention some unclear locations, such as outdoors, or a city), 2 (mention some specific locations, such as home, office, or meeting room). 
    \item People: 0 (not mention and impossible to infer people involved), 1 (mention some unclear people information), 2 (mention some people clearly, such as parents, girlfriend, or myself).
    \item Emotion: 0 (not mention at all), 1 (mention a general feeling), 2 (describe the reasons and thoughts related to the emotion mentioned).
    \item Activity: 0 (unable to remember), 1 (mention what happened without details), 2 (detailed description of what happened).
\end{itemize}

Finally, they scored the remaining recollections under this criteria, and resolved all the disagreements further.

\textbf{System Usability.} There is always a trade-off between functionality and usability in systems \cite{goodwin1987functionality}.
We referred to the standard System Usability Scale (SUS) tool to assess the system usability in two aspects \cite{brooke1996sus}: 1) easy to learn; 2) easy to use.

The above mentioned measurements were designed for both the Baseline system and \tool. Meanwhile, there were also additional three questions about \tool\!. Specifically, 1) whether they were satisfied with \tool\!'s prediction (\textit{satisfaction}); 2) whether they checked \tool\!'s prediction carefully (\textit{carefulness}); 3) whether they had privacy concerns (\textit{privacy concern}).   

All questions in the questionnaire were measured with a 5-point Likert scale, with 1 being the most negative impression (e.g., strongly disagree) and 5 being the most positive impression (e.g., strongly agree).
\section{Results}\label{sec:results}
\begin{figure}[!ht]
    \includegraphics[width=8cm]{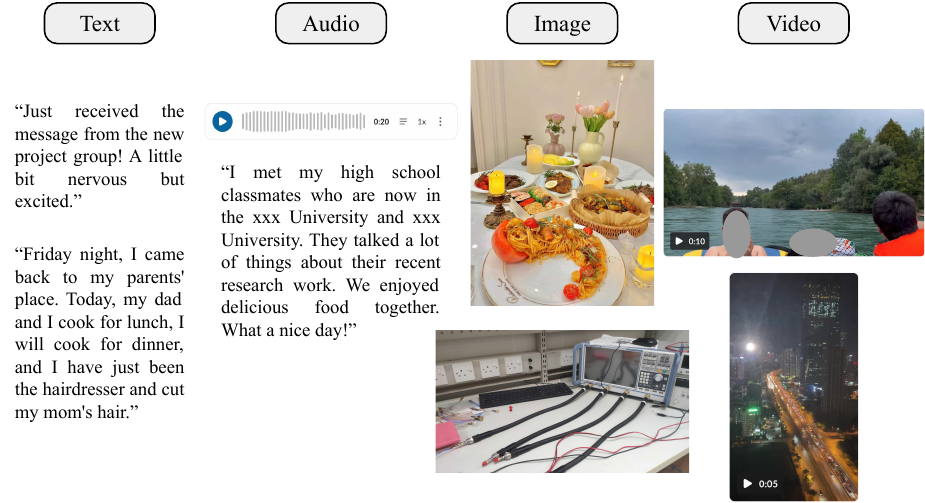}
    \caption{Samples of recorded diaries in different modalities.}
    \label{fig:samples}
    \Description{Figure 3 displays four modalities of samples from recorded diaries: text, audio, image, and video. In the 'Text' category, there are two examples of participants' recordings of expressing excitement about transitioning to a new project group and the happiness of visiting parents. The 'Audio' category includes a transcript of a voice message discussing a meeting with high school classmates. In the 'Image' section, there are photographs showcasing exquisite food and laboratory equipment. Lastly, the 'Video' section features screenshots from videos, one depicting a kayaking activity and another showing the nightscape of a city center.}
    \centering
\end{figure}

We represented some samples of the recorded diaries in different modalities in Figure \ref{fig:samples}. In the following sections, we analyzed participants' recording behaviors, interactions with \tool\!, and contents provided in the elicitation interview based on the measurements described in Section \ref{sec:exp_eval}.

\subsection{Recording Habits}

Towards the collected diaries in the recording period, we investigated participants' choice of modality to record diaries and their recording timing during the day respectively. 
\rr{We also measured the amount of contextual information contained in the recorded diary entries.}
Based on the comparison between the Baseline system and \tool\!, we found that \tool increased participants' willingness to take photos as diaries and revealed empathy, especially during the recording of negative emotions.

\subsubsection{Choice of Modality}\label{sec:modality}

In the recording period, 12 participants recorded 105 diary entries (Mean=8.75, S.D.=4.38) during the week using the Baseline system, and recorded 104 diary entries (Mean=8.67, S.D.=5.89) during the week using \tool\!. 
They recorded diary entries in different modalities based on their personal preference or the surroundings. 
The number of captured diaries with different modalities using the Baseline system and \tool are shown in Table \ref{tab:modality}. 
The recorded diary including both the text and image was considered as a recording of hybrid modality.

\textbf{Text and Image} Generally speaking, we observed that text was the most often adopted modality, and image was also widely considered. 
Some participants thought that typing a piece of text is efficient to express themselves, and convenient for editing. 
For example, P12 mentioned that ``\textit{It is a natural way to record something by text and language. There is some room for thought when writing it out. Even if you made a typo, or suddenly don’t want to continue editing the current sentence, you can delete it at any time before sending it out}''. 
Participants' choice of modality also depended on their emotion and the type of the recorded events. 
For instance, P6 said that ``\textit{I usually feel that when my pressure is low, I subconsciously like to use photos to record.
But when I feel that I am more annoyed and busy, I usually use text to record}''. 
And P7 mentioned that ``\textit{When I record some events in my spare time, I might prefer using photos, but when it comes to work, there are more text}''.

\textbf{Audio and Video}
\rr{We found that participants were not used to record audios and videos during the recording period as shown in Table \ref{tab:modality}.}
However, P11 recorded 3 audio clips when using \tool\!. 
P11 said that ``\textit{The main reason for recording audio is because it’s not easy for me to type when I’m walking or taking the subway, so I just speak and then \tool could help me enrich the description of the event}''. 
P1 and P6 recorded 2 short video clips (lasting 10 seconds and 5 seconds respectively). 
Participants indicated that taking photos is enough for them to record an event clearly comparing to recording videos.

\textbf{\tool\!'s Impact on Choice of Modality} By observing the Table \ref{tab:modality}, there is an obvious increase in the number of diaries recorded by image, and a decrease of both the recording with text and hybrid one (text\&image). 
The reason is participants tended to rely on \tool to understand the image and generate the relevant contextual information, which lightened their burden of recording. 
For example, P5 described this change of recording habit, ``\textit{In the second week when I used \tool\!, I felt taking photos may be more convenient, and \tool can directly help me generate some textual supplements without typing by myself}''.

\begin{table}[!ht]
    \caption{The number of diaries recorded by each modality.}
    \label{tab:modality}
    \centering
    \begin{tabular}{ccc}
    \toprule
       Modality  & Baseline & \tool \\
    \midrule
       Text & 62 & 50 \\
       Image & 11 & 35 \\
       Text \& Image & 30 & 16 \\
       Audio & 0 & 3 \\
       Video & 2 & 0 \\
    \midrule
    Total & 105 & 104 \\
    \bottomrule
    \end{tabular}
\end{table}

\subsubsection{Timing of Recording}\label{sec:time_record}

\begin{figure}[ht]
    \includegraphics[width=8cm]{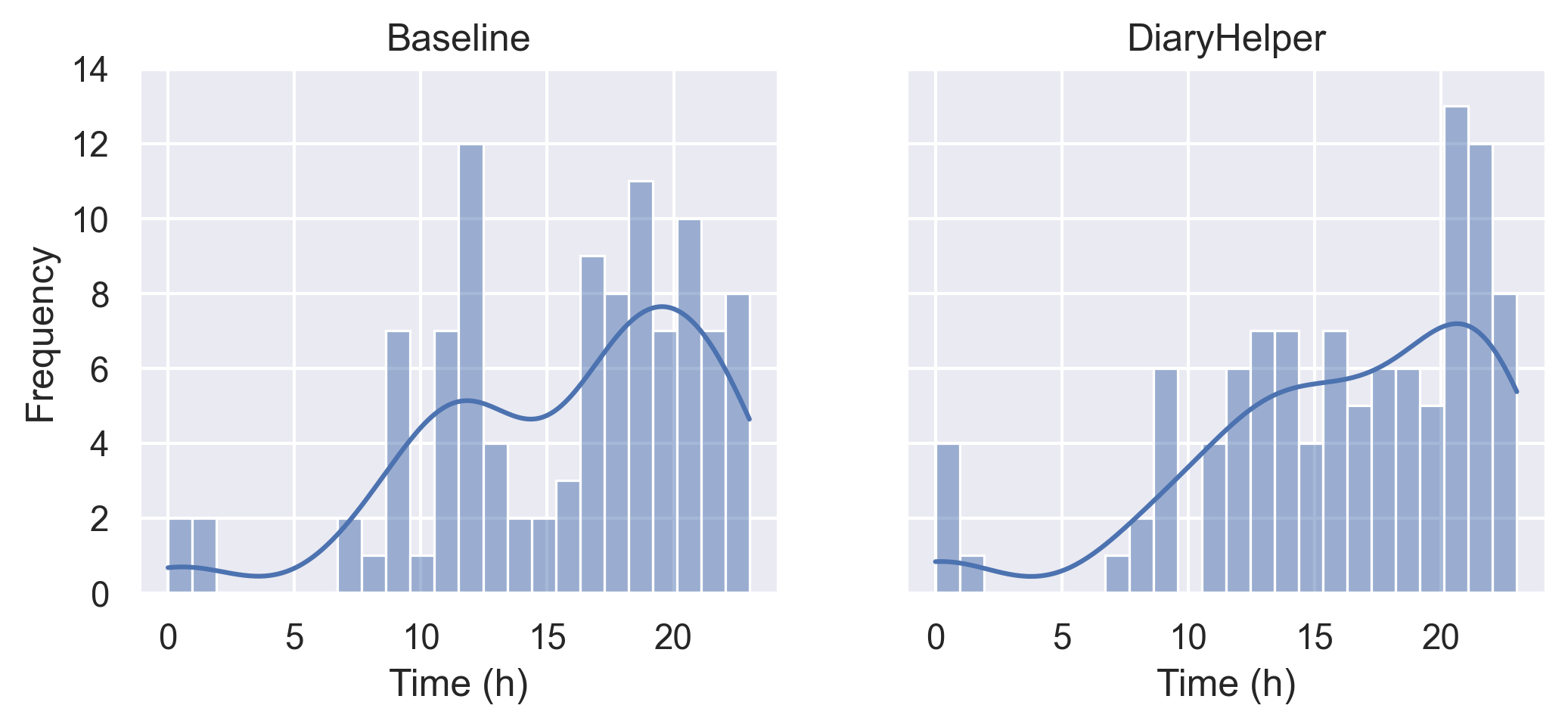}
    \caption{The number of participants' recorded diary entries per hour during a day.}
    \label{fig:time_dist}
    \Description{Figure 4 visualizes the number of diary entries in each hour during a day for both the Baseline system and \tool in two histograms. Both sub-figures display data across a 24-hour period. In the 'Baseline' chart, frequency peaks between 11:00 and 15:00 , with the highest frequency of 12. The \tool chart shows a continuous increase of frequency through a day, where there is a sharp peak at a frequency of 13 after 20:00, indicating participants' preference of recording during the night.}
    \centering
\end{figure}

For exploring when typically participants recorded diaries, we visualized the number of diary entries in each hour for both the Baseline system and \tool in Figure \ref{fig:time_dist}.
During the daytime, there was not a significant difference in the recording frequency between these two systems. 
One reason is some participants had very busy working schedules during the recording period and often had to work overtime to midnight. In such cases, they thought \tool could provide emotional support for them by expressing the mood, and would like to record this event with higher willingness. 
For example, P4 mentioned that ``\textit{\tool let me feel like a friend waiting for me. When I was tired after working overtime, I would like to talk to \tool and express my feelings. I feel like I am empathized when I notice a negative emotion tag in the memo}''.

\rr{\subsubsection{Amount of Contextual Information }\label{sec:aci_test}
To measure the amount of contextual information contained in the recorded diary entries, we adopted the same measurements as recall evaluation to score five dimensions of contextual information mentioned in Section \ref{sec:exp_eval}.
Due to the indirect contextual information embodied in the captured images and videos, we only included diaries entries of `Text', `Text \& Image' and `Audio' modalities.
For the audio-based diary entries, we labeled the corresponding transcriptions.
Two authors manually labeled 92 diary entries from the Baseline system and 69 diary entries from \tool\!.
We set the hypothesis that the amount of contextual information contained in the diary entries under the two systems were different in five dimensions (i.e., time,
location, emotion, people and activity). 
Since the score for each dimension was ordinal (0,1 or 2), we applied Wilcoxon rank-sum
test \cite{wilcoxon1970critical} to measure the statistical significance.
We reported the test statistic, p-value, and effect size \footnote{\rr{The effect size is used to indicate the strength of relationship between the two group of data. It equals to the difference between the mean of two groups divided by the pooled standard deviation of two groups. Different statistical test may have different criteria to interpret the amount of effect size.} For Wilcoxon rank-sum test, effect size is commonly interpreted in three levels with different ranges: 0.10 - < 0.3 (small effect), 0.30 - < 0.5 (moderate effect) and >= 0.5 (large effect) \cite{rstatix}}
for each dimension in Table \ref{tab:hypo_test_info_amount}.
Meanwhile, we conducted another two hypothesis tests for the Baseline system and \tool respectively to investigate whether the carryover effect existed.
For these two tests, we set the hypothesis that the amount of contextual information contained in diary entries between two groups when using each system were different in five dimensions.
The corresponding results are shown in Table \ref{tab:hypo_test_info_amount_co_b} and \ref{tab:hypo_test_info_amount_co_d}.

From Table \ref{tab:hypo_test_info_amount}, we found that participants recorded similar total amount of contextual information ($W=1.83, 0.05<p<0.1$).
For each specific dimension, only the contextual information of time revealed a difference ($W=3.11, p=0.002$), which indicated participants tended to record more time-related information when using the Baseline system.
Regarding the results of checking the carryover effect, we did not observe the carryover effect on the total amount of contextual information recorded between the two groups for both the Baseline system ($W=1.52, p>0.1$) and \tool ($W=1.88, 0.05<p<0.1$). It indicated that participants recorded the similar amount of contextual information without the effect of order of usage between the two systems.
}

\subsubsection{Questionnaire Feedback.}\label{sec:record_ques} As mentioned in Section \ref{sec:exp_eval}, we assessed the participants' recording habits in three dimensions and system usability in two dimensions. Participants' feedback in the questionnaire is shown Table. \ref{tab:result_questionnaire}.

We observed that participants showed a higher willingness to record events frequently and immediately using \tool compared to the Baseline system.
And their perception of \textit{modality} was consistent with our analysis in Section \ref{sec:modality}, which demonstrated participants would like to adopt diverse modalities to record diaries for adapting to different situations.
\rr{Even though an additional function of checking and submitting memo was integrated into \tool\!, participants did not show difficulties of learning and using \tool\! based on their feedback regarding usability.}

\begin{table}[!ht]
    \caption{Participants' feedback of the questions regarding the recording, elicitation and usability.}
    \label{tab:result_questionnaire}
    \centering
\begin{tabular}{cccc} 
\toprule
\multirow{2}{*}{Category} & \multirow{2}{*}{Factor} & Baseline & \tool  \\ 
 &  & Mean(S.D.) & Mean(S.D.) \\ 
\midrule
\multirow{3}{*}{Recording} & Frequency & 2.83(0.80) & 4.00(0.58) \\
 & Timeliness & 2.25(0.60) & 3.25(0.72) \\
 & Modalities & 2.25(0.43) & 3.92(0.95) \\ 
\midrule
\multirow{3}{*}{Elicitation} & Recall & 2.75(0.83) & 4.42(0.49) \\
 & Relation & 3.17(1.07) & 3.58(1.04) \\
 & Reflection & 3.67(0.85) & 4.33(0.62) \\  
\midrule
\multirow{2}{*}{Usability} & Easy to learn & 4.42(0.64) & 4.58(0.49) \\
 & Easy to use & 4.33(0.75) & 4.33(0.75) \\
\bottomrule
\end{tabular}
\end{table}

\subsection{Performance of \tool}

\tool is empowered by GPT-3.5 and cloud service APIs introduced in Section \ref{sec:implementation}. 
It is crucial to evaluate whether the predicted values of contextual information are proper and accurate without too much modification in the memo generated by \tool\!. 
Since the time section in memo was filled in automatically based on the metadata of the message, we mainly focused on the evaluation of other four dimensions of contextual information.
\rr{
Most of the participants (9/12) checked and submitted all the generated memos from \tool\!. 
The other three participants received 31 generated memos in total, and they checked and submitted 22 memos accordingly.
In this section, we only included the submitted memos for the performance evaluation, and considered confirmed contextual information by participants as the ground-truth.
}
We observed that \tool showed the best prediction ability on people, and the prediction of location, emotion and activity were similar and decent.

\subsubsection{Emotion}

We considered the selected sentiment label by participants as the ground-truth label.
Then we compared the predicted label and ground-truth label using classic metrics in sentiment analysis task.
\tool achieved 0.68 average accuracy and 0.69 macro average F1-score, which demonstrated \tool could understand participants' emotional states in most cases. 
The detailed prediction results for each sentiment label are shown in Table \ref{tab:sentiment}. 
For the diaries with a negative sentiment, \tool\!'s prediction achieved the highest F1-score, which means \tool could empathize with participants’ negative emotions.

\begin{table}[ht!]

    \caption{The performance of emotion prediction of \tool\!.}
    \label{tab:sentiment}
    \centering
    \begin{tabular}{ccccc}
    \toprule
    Sentiment & Precision & Recall & F1-score & Proportion\\
    \midrule
    Positive & 0.80 & 0.43 & 0.56 & 43.07\% \\
    Neutral & 0.61 & 0.88 & 0.72 & 44.62\% \\
    Negative & 0.76 & 0.81 & 0.79 & 12.31\%\\
    \bottomrule
    \end{tabular}
\end{table}

\subsubsection{Location, People and Activity}\label{sec:multiple}

For these three dimensions of contextual information, participants could choose multiple possible options from \tool\!'s prediction. 
\tool selected the most possible option automatically before confirming with participants. 
If this pre-selected option is included in the participants' final choices, it is counted as a hit. 
Among all the submitted memos, we calculated the hit ratio, and the proportion of submissions with different number of selected options, which were shown in Table \ref{tab:submission_evaluation}. 

\textbf{Location} The prediction of location achieved 0.59 hit ratio, and participants also selected another possible location tag in 6.92\% of submissions. 
When participants could find some reasonable location tags, they still added some additional location information in 3.33\% of submissions.
However, among 30.77\% of submissions, participants could not find an option to describe the location properly. 
In such cases, participants would like to type in the location information in 27.5\% of submissions.
Generally, the prediction of location performed well for most diaries. After checking the prediction with errors, we found that \tool could not understand the diaries clearly in these cases: 1) screenshots, 2) close-up photos of special objects (\eg movie ticket or leaflet), 3) text or audio based description without clear location information.

\textbf{People} Among 90\% of submissions, participants accepted the predicted people tags, which indicated an excellent understanding ability of people involved in the recorded diary. 
Participants selected another people tag in 3.85\% of submissions to enrich the description of people.

\textbf{Activity} The hit ratio of activity prediction was 0.60, which was similar to that of location prediction.
In 23.85\% of submissions, participants could not find a proper description of the event happened. 
However, in this case, participants added some additional notes in 80.65\% of submissions, which indicated their high motivation to complete the description of activities when \tool can not do well.
If the generated activity descriptions make sense, participants still left some additional notes in 25.25\% of submissions. 
We also observed that the predicted contextual information in the memo could inspire participants to annotate more relevant information. 
For example, \tool predicted one potential activity option ``\textit{attending violin lesson after work}'' for P12's one recorded diary. 
Then P12 selected this option and expanded it into ``\textit{I started playing violin since July, and I tried to practice every day. I enjoy playing it}'' as the additional note.
Overall, \tool\!'s prediction ability on activity was adequate and assist participants to describe the events of interest efficiently.

\begin{table}[ht!]

    \caption{The prediction performance for the contextual information with multiple choices.}
    \label{tab:submission_evaluation}
    \centering
    \begin{tabular}{cccccc}
    \toprule
    \multirow{2}{*}{Category} & Hit Ratio & \multicolumn{4}{c}{Prop. of \# Selected Options} \\
    \cline{3-6}
     & Mean(S.D) & 0 & 1 & 2 & >2 \\
    \midrule
    Location & 0.59(0.17) & 30.77\% & 62.31\% & 6.92\% & - \\
    People & 0.90(0.11) & 0 & 96.15\% & 3.85\% & 0 \\
    Activity & 0.60(0.19) & 23.85\% & 55.38\% & 12.31\% & 8.46\% \\
    \bottomrule
    \end{tabular}
\end{table}

\subsubsection{Questionnaire Feedback.} 
\begin{figure}[ht!]
    \includegraphics[width=8.5cm]{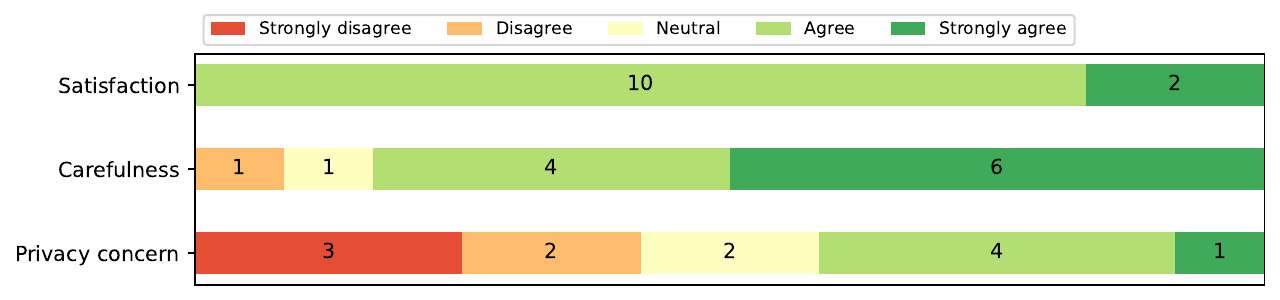}
    \caption[]{Participants' ratings of \tool regarding satisfaction, carefulness and privacy concern.}
    \label{fig:last3questions}
    \Description{Figure 5 is a stacked bar chart displaying participants' ratings of \tool across three categories: Satisfaction, Carefulness, and Privacy Concern. Each category has five response options: Strongly Disagree, Disagree, Neutral, Agree, and Strongly Agree, each represented by a different color. For Satisfaction, the majority of participants (10) selected 'Agree', and 2 participants chose 'Strongly Agree', indicating a positive response. For Carefulness, 1 participant selected 'Disagree', 1 chose 'Neutral', 4 selected 'Agree', and the majority, 6 participants, chose 'Strongly Agree', suggesting a generally careful checking of the memos. Regarding Privacy Concern, 3 participants 'Strongly Disagree', 2 chose 'Disagree', 2 were 'Neutral', 4 selected 'Agree', and 1 chose 'Strongly Agree', showing an overall neutral attitude on privacy concerns.}
    \centering
\end{figure}

We also asked participants three additional questions about the interaction with \tool specifically as mentioned in Section \ref{sec:exp_eval}, and their ratings were shown in Figure \ref{fig:last3questions}.
We observed that all participants were satisfied with the performance of \tool (10 agree and 2 strongly agree), and most of participants checked the generated contextual information carefully (4 agree and 6 strongly agree). 
However, P8 mentioned that ``\textit{After I found that the contents generated by \tool were good, I tended to check the information less carefully than before}''.
Meanwhile, part of the participants had privacy concerns about the posted data (4 agree and 1 strongly agree).
Therefore, in the onboarding phase of the study, we should emphasize how we will protect participants' personal data, and remind participants to record data without confidential or sensitive information.

\subsection{Quality of Elicitation Interview}
After analyzing participants' recording habits and system performance of \tool\!, we explored the richness of participants' recall and provided insights during the elicitation interview in this section. 
We found that participants could recall more details for each recorded diary entry and provide more insights related to the study topic under the assistance of \tool\!.

\subsubsection{Evaluation on Recall}\label{sec:eval_recall}

Two authors scored participants' recall for 209 diary entries (105 diaries recorded by the Baseline system and 104 diaries by \tool\!).
\rr{We followed the same procedure and type of hypothesis test mentioned in Section \ref{sec:aci_test} to evaluate participants' recall under the two systems. 
We set the hypothesis that the usage of \tool could trigger participants to enrich their recall in five dimensions (i.e., time, location, emotion, people and activity).
Then we reported the test statistic, p-value, and effect size for each dimension in Table \ref{tab:hypo_test}.
We also carried out another two hypothesis tests to investigate whether the carryover effect existed.
For these two tests, we set the hypothesis that the richness of recall provided by the participants from two groups under each system was different in five dimensions.
The corresponding results are shown in Table
\ref{tab:co_recall_b} and \ref{tab:co_recall_d}.}


\begin{table*}[ht!]
    \caption{
    The statistical analysis of recall evaluation for the Baseline system and \tool using the Wilcoxon rank-sum test, where the p-value (-: p > .100, +: .050 < p < .100, *:p < .050, **:p < .010, ***:p < .001) is reported. Effect size with large or moderate magnitude is \textbf{bolded}.}
    \label{tab:hypo_test}
    \centering  
\begin{tabular}{cccccccc} 
\toprule
\multirow{2}{*}{Category} & Baseline & \tool & \multicolumn{4}{c}{Statistics} & \multirow{2}{*}{Hypothesis} \\ 
\cline{4-7}
 & Mean(S.D.) & Mean(S.D.) & W & p-value & Sig. & Eff. Size & \\ 
\midrule
Time & 1.05(0.65) & 1.04(0.59) & 0.11 & 0.909 & - & 0.01 & Rej. \\
Location & 0.81(0.83) & 1.39(0.76) & 4.64 & <0.001 & *** & \textbf{0.73} & Acc. \\
Emotion & 0.75(0.83) & 1.07(0.89) & 2.39 & 0.017 & * & \textbf{0.37} & Acc. \\
People & 1.02(0.73) & 1.40(0.67) & 3.53 & <0.001 & *** & \textbf{0.55} & Acc. \\
Activity & 1.57(0.49) & 1.73(0.44) & 1.99 & 0.047 & * & \textbf{0.34} & Acc. \\
\midrule
Total & 5.20(2.05) & 6.63(1.71) & 5.00 & <0.001 & *** & \textbf{0.76} & Acc. \\
\bottomrule
\end{tabular}
\end{table*}

From Table \ref{tab:hypo_test}, we observed that hypothesis regarding time was rejected, which means participants could recall similar temporal information using the Baseline and \tool\!. 
One possible reason is that participants tended to record an event immediately when they thought an event was interesting to record (Section \ref{sec:record_ques}), so the captured temporal information by \tool was similar to that captured in the Baseline system.

The richness of recall regarding location ($W = 4.64, p < 0.001$) and people ($W = 3.53, p < 0.001$) increased significantly when using \tool\!, compared to the Baseline system (same significant level with large effect).
The richness of remembered emotion ($W = 2.39, p = 0.017$) and activity ($W = 1.99, p < 0.047$) also increased significantly, but the significant level and effect size were lower than those of location and people information.
It is worth mentioning that the score regards emotion information raised from a relatively low score (<0.8) to a notable coverage (1.07).
Overall, the total score of each recall represented a significant increase, which indicated the effectiveness of \tool to trigger participants' detailed recall based on the auto-generated, structured, comprehensive contextual information. 

Meanwhile, some generated contents in the memo also reminded participants of noting down other valuable contextual information which they might be forgotten later. 
For instance, P6 described his change of attitude on contextual information:
\begin{quote}
    ``\textit{If let me to record diaries only by myself, I might not write down such detailed contextual information, because I generally feel that these information might not be very helpful for my recall. However, it turns out that these information captured by \tool are still valuable to my memories.}''
\end{quote}

And P7 mentioned that ``\textit{I can find the key points I need at a glance of the memo, including location, activities and so on. This really allows me to recall events very quickly and easily, and to remember more details}''.

\rr{
Based on the results in Table \ref{tab:co_recall_b}, we observed the carryover effect of the total score of recall when using the Baseline system ($W=2.04, p=0.041$). 
However, the corresponding p-value was close to $0.05$, which indicated the significance level to accept the hypothesis was relatively low.
From Table \ref{tab:co_recall_d}, we did not observe the carryover effect of the total score of recall for the \tool system ($W=1.69, 0.05 < p < 0.1$), but there was a minor carryover effect shown for recall of location ($W=2.40, p=0.016$).}

\balance

\subsubsection{Evaluation on Reflection} Participants were asked about their general feelings of work-life balance and strategies to manage work and life after recalling the recorded diaries as described in Section \ref{sec:exp_pip}. 
First, we found that high quality of participants' recall was crucial for discussion about the study topic. 
For instance, P2 mentioned that ``\textit{If I could have a more complete recall of what happened at that time, such as my emotions, or who I was doing it with, it would help me better reflect on whether I had a good work-life balance in the past week}''.
Second, emotion as one dimension of contextual information captured by \tool was useful for participants to reflect on their mental state and adjust their strategy to balance work and life.
For example, P4 mentioned that ``\textit{After having the emotional tag, I can clearly grasp the changes in my emotions this week, allowing me to further adjust the rhythm of my work}''.
Third, participants thought it was time-consuming to review long pieces of text and audio clips lasting a long time in the elicitation interview. In such cases, they tended to check the concise and structured memo summaries (Figure \ref{fig:interface}, \textcircled{5}) generated by \tool directly, which could facilitate participants' holistic understanding and deep reflection of their status and daily behaviors.
For instance, P7 mentioned that ``\textit{After reviewing all the summaries, it makes me better understand my arrangements regarding work and life}''.

\subsubsection{Questionnaire Feedback.} Participants also rated system's assistance regarding \textit{recall}, \textit{relation} and \textit{reflection} in the elicitation interview mentioned in Section \ref{sec:exp_eval}.
Based on Table \ref{tab:result_questionnaire}, we found that the average rating of \tool\!'s assistance for \textit{recall} increased the most (from 2.75 to 4.42), compared to the Baseline system. 
This result is consistent with the statistical test results in Section \ref{sec:eval_recall}.
\tool\!'s assistance for \textit{reflection} (4.33) was also better than that of the Baseline system (3.67).
However, system's assistance for \textit{relation} is similar for the Baseline system (3.17) and \tool (3.58).
P12 mentioned a possible reason: ``\textit{Since there is no contextual information in the memo designed explicitly for this study topic, I do not find the effect of \tool to help me realize the relation between my recorded diary and study topic}''.





\section{Discussion}
Participants involved in elicitation diary studies face challenges of decreasing willingness to record diaries, postponing the recording due to the time limits, and feeling perplexed regarding how to record an event of interest clearly \cite{Carter2005media,clark2004framing,Chong2015cuenow,Brandt2007snippet,Gorm2017photo}.
Our work demonstrates a promising approach of designing an automatic contextual information recording agent to assist participants in recording diaries and trigger them to recall more details and discuss informatively in the elicitation interview. 
\rr{The rest of this section presents the deeper insights of \tool for future implementation of similar tools and some limitations.}


\subsection{Usefulness of Automatic Contextual Information Recording}
Prior research has emphasized the importance of minimizing the recording burden in diary studies across various contexts \cite{Brandt2007snippet,Carter2005media,bylsma2011crying}. 
Results from system testing (Table \ref{tab:sentiment}, \ref{tab:submission_evaluation}) and participant feedback (Table \ref{tab:result_questionnaire}, Figure \ref{fig:last3questions}) indicate that \tool achieves high accuracy in contextual information generation and elicits high levels of user satisfaction. 
Participants appreciated the low recording burden of \tool\!, which allowed for quick and efficient recording even during fragmented periods of time. 
These findings suggest that automatic generation of contextual information can effectively reduce recording burden in applications that require rapid data collection. 

As mentioned in Section \ref{sec:multiple}, the predicted options of contextual information provided by \tool can serve as prompts for participants to annotate more relevant information in their diaries. 
This emphasizes the importance of providing auto-generated information as prompts to assist participants in understanding what and how to record. 

The five dimensions of contextual information generated by \tool are general memory cues that have been demonstrated to be useful for episodic memory \cite{barsalou_1988,Gouveia2013footprint}.
However, for specific diary study topics, customized cues could be integrated into the memo to better support participants' recording needs. 
For example, P9 suggested that ``\textit{A tag indicating my current work-life relationship may also be helpful for me to keep my thoughts at that time, and a visualization of my proportion of work and life could also be represented}''. 
\rr{In the future design of agents like \tool\!, researchers may consider the integration of various forms of visualizations to summarize recorded diaries and contextual information across diverse dimensions \cite{Katerina2007diaryvis,Myriam2013emodiary,van2006interviewviz}.
For instance, under our study topic about work-life balance, we can consider design an interface to show the contextual information integrally and dynamically, such as the proportions of work and life related events, the changes of emotions, and the clustered location information.
It could increase participants' understanding of past events in detail and assist them in forming holistic view regarding the study topic potentially.
However, the timing of presenting visualizations needs further study to avoid intervening participants' own perception and justification. 
}

At present, \tool generates a fixed amount of contextual information throughout the recording period. 
However, a more sophisticated approach to contextual information generation could be incorporated in future designs. 
For instance, at the beginning of the study, participants may be less familiar with the recording process, and thus, more extensive contextual information could be provided to facilitate their engagement and comprehension. 
As the experiment progresses and participants become more experienced with the recording process, the amount of contextual information provided could be gradually reduced to avoid overloading participants with redundant or irrelevant information. 

\subsection{Empathy and Interactivity}
Maintaining user interest and engagement is a critical factor in the success of diary studies \cite{anjeli2013guide}. 
While \tool cannot engage in casual conversations with participants, its prompt acknowledgment and interactions with the memo have shown great potential to elicit a sense of empathy and increase participants' willingness to record their diaries. Participants perceived \tool as a companion in the Slack channel, and considered recording their diaries as sharing with a friend. 
For example, P12 noted that ``\textit{When some events of interest happened in my life, I could remember to share it with \tool\!, which improves my motivation to record diaries}''. 
This sense of companionship and support provided by \tool was particularly important in cases where participants experienced negative emotions, as they could feel empathized when noticing an emotion tag in the memo generated by \tool\! (as mentioned in Section \ref{sec:time_record}).

\rr{In prior diary studies, researchers typically played the role of acknowledging participants' recordings and giving feedback through emails or text messages \cite{Gorm2017photo,vega2023gig}.}
However, \tool represents a potentially valuable approach to enhancing participants' willingness to record diaries and fostering empathetic feelings. 
\rr{It is particularly important when long-lasting diary studies may suffer from participants' decreased motivation (or even fatigue) in recording \cite{Gorm2017photo}.}
\rr{It is a promising direction for future work to investigate how adopting a chatbot into the recording platform may help provide a more personalized and interactive experience for participants, and maintain their interest and motivation over time.}

\subsection{Generalizability}
The potential generalizability of \tool is a promising development for other research methods that rely on participants' self-recorded data. 
For instance, experience sampling methods (ESMs) involve participants reporting on their experiences in real-time, such as their thoughts, feelings, and behaviors \cite{csikszentmihalyi2014experience,csikszentmihalyi2014validity,hektner2007experience,kang2022emotionesm}. 
\tool could potentially be adapted to ESM studies to automatically generate prompts or cues based on participants' self-recorded data, thereby improving the quantity, accuracy, and completeness of the data collected. 
Zhen \etal's work has highlighted the potential of integrating photos into ESM studies \cite{zhen2014photoesm}, where \tool could consider participants' sampled experiences and photos together to infer richer contextual information. 
Moreover, photovoice methods involve participants representing their experiences through photography followed by a discussion session \cite{wang1997photovoice,catalani2010photovoicereview,sutton2014photovoice}.
The photo-taking process and discussion could be leveraged by \tool\!'s abilities in contextual information extraction and summarization.

Overall, by automating the generation of relevant prompts and cues, intelligent agents like \tool can enhance the quality and quantity of self-recorded data collected in a range of research methods, thus enabling more robust and insightful analyses. 

\subsection{Limitations and Future Work}
Our system design and experiment have several limitations.
First, we only conducted one diary study in our experiment, as the five dimensions of contextual information recorded by \tool are not specific to any particular diary study and can be implemented flexibly across different topics. 
However, future studies could explore the differences in participants' interaction patterns and issues across diverse diary studies. 
Second, the topic selected for our diary study focused on ``work-life balance'', which encompasses a wide range of relevant events and may have made it easier for participants to identify and record events. 
For diary studies targeting specific situations or groups, such as cyber security of smart home devices \cite{sarah2022cybersecurity} or gig economy freelancers \cite{vega2023gig}, it would be valuable to investigate how \tool affects participants' recording habits and discussions.
Third, our recruited participants were young adults (aged 22-28), which may introduce some bias in our experiment results.
Future studies should aim to recruit participants from diverse age groups to comprehensively evaluate the performance of \tool\!. 
Additionally, alternative designs for individuals who are unfamiliar with electronic devices should also be considered.
Fourth, more comparison experiments with varying recording periods could be conducted to explore the effectiveness of the memory cues recorded by \tool for events that occurred at different times in the past. 
In particular, for studies requiring long-term observations, it would be valuable to analyze the impact of contextual information captured by \tool on participants' recall after a long time.
Moreover, the diary entries as memory anchors could help ground participants' recall, but it may also steer their recollection away from details not recorded in the diary.
In our study, we just followed standard practices to let participants access the recorded diaries during the recall session.
Finally, while Carter \etal's work supports the use of tangible objects as a modality for diary recording \cite{Carter2005media}, \tool is not equipped to handle cases where participants keep physical objects as data entries. 
One possible solution is to encourage participants to take a photo of the object and submit it to \tool as a recording. 

\rr{In the future, we plan to invite some domain experts in diary studies to discuss and evaluate the feasibility and potential issue of \tool\!. 
Meanwhile, besides the recording burden and recall focused in this work, we will continue to explore other factors that may affect participants' insight and reflection, such as adopting AI-summarization of diary behaviors for elicitation.}



\section{Conclusion}
In this work, we designed \tool\!, an automatic contextual information recording agent for elicitation diary study.
\tool is empowered by cloud service APIs to process multimodal diary data and LLM to understand and generate contextual information.
Compared to the Baseline system, a within-subject diary study demonstrated that \tool improved participants' willingness to diary recording and decreased the burden of recording contextual information in the recording period.
For the elicitation interview, \tool triggered participants to recall captured diaries with more detail and provide more insights regarding the study topic.
We further summarized the design implications to guide the future approach proposed to enhance the diary study methods.

\begin{acks}
This work is supported by the Research Grants Council of the Hong Kong Special Administrative Region under General Research Fund (GRF) with Grant No. 16203421.
We thank anonymous reviewers for their valuable feedback and our study participants for their great contribution and effort. 
Lastly, we appreciate Yubo Xie and \href{https://orcid.org/0000-0001-8199-071X}{Zeyu Huang}'s insightful discussions. 
\end{acks}

\bibliographystyle{ACM-Reference-Format}
\bibliography{references}

\appendix
\section{Prompt Used for GPT model}\label{app:prompt}
Here is an exact prompt used to generate contextual information for image-based diaries:

``You are an experienced diary study researcher. 
You are conducting a diary study right now, and when you receive the data captured by the participant, you need to help the participant to record some contextual information. 
These contextual information will be used as the cues for the participant to recall the event. 
In this way, we could collect more useful and abundant information from the participant in the interview after the logging period. 
Now, in the logging period, one participant capture one image as one diary entry. 
The objects detected in this image are \textit{[list of objects]} (ranked by the decreasing order of confidence).
The description of this image is \textit{[textual description]}.
Please predict the following contextual information based on the aforementioned information:

Location: predict three possible point of interest locations, you could use the point of interest location categories in Google Maps or some other location-based service apps.

Emotion: select only one from these three categories, Positive, Neutral and Negative, please keep the same spelling.

People: select only one from these five categories, Alone, Families, Friends, Colleagues and Acquaintances, please keep the same spelling.

Activity: give six descriptions of the six possible activities in this scenario (give more details for each activity, but each description should be less than 151 characters).

Finally please output these information in English in valid JSON format. And the value for the Location and Activity should be a list of three and six elements respectively.

\textit{EXAMPLE: \{"Location": [Library, Workspace, Meeting room], "Emotion": Positive, "People": Colleague, "Activity": [Working on laptop and taking notes, Studying or doing research, Planning or organizing tasks for the day, Preparing a meeting, Watching a academic seminar, Discussing the current project]\}}''

\section{Figures and Tables}
\begin{figure}[ht!]
    \includegraphics[width=7cm]{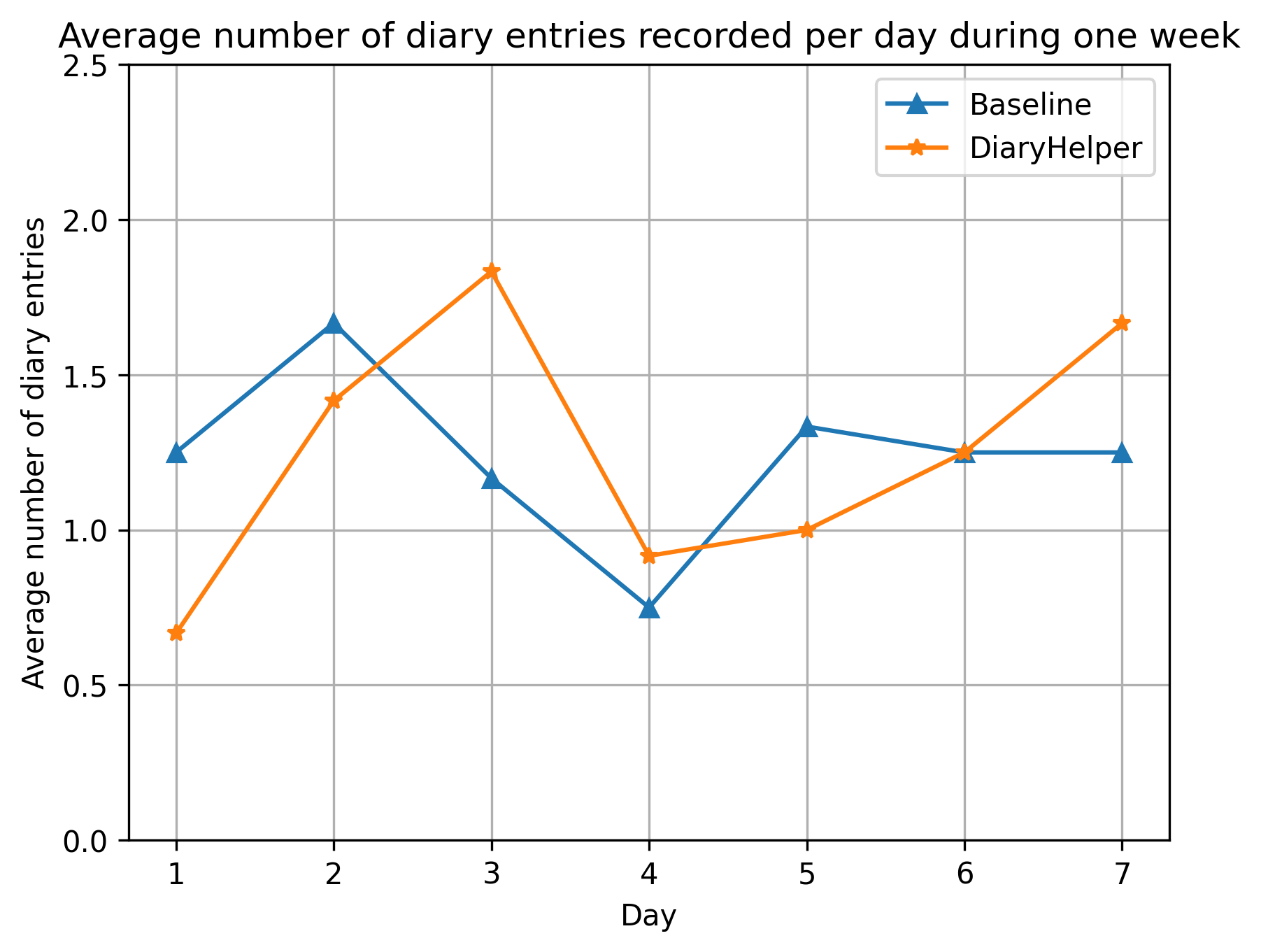}
    \caption{The average recordings per day during one week.}
    \label{fig:avg_record}
    \Description{Figure 6 displays a line chart showing the average number of diary entries per day over a week. Two lines represent the Baseline system and \tool\!. The `Baseline' fluctuates throughout the week, starting around 1.5 entries, peaking on Day 2, and ending just below 1.5 entries on Day 7. The \tool starts at around 0.6 entry, peaks at around 1.7 entries on Day 3, and finishes the week above 1.5 entries. The chart illustrates the average number of recorded diaries are similar under the two systems through one week.}
    \centering
\end{figure}

We present Figure \ref{fig:avg_record} and Table \ref{tab:hypo_test_info_amount}-\ref{tab:co_recall_d} to support the result analysis in Section \ref{sec:results}.

\begin{table*}[b]
    \caption{
    The statistical analysis of the amount of contextual information contained in diary entries for the Baseline system and \tool using the Wilcoxon rank-sum test, where the p-value (-: p > .100, +: .050 < p < .100, *:p < .050, **:p < .010, ***:p < .001) is reported. Effect size with large or moderate magnitude is \textbf{bolded}.}
    \label{tab:hypo_test_info_amount}
    \centering  
\begin{tabular}{cccccccc} 
\toprule
\multirow{2}{*}{Category} & Baseline & \tool & \multicolumn{4}{c}{Statistics} & \multirow{2}{*}{Hypothesis} \\ 
\cline{4-7}
 & Mean(S.D.) & Mean(S.D.) & W & p-value & Sig. & Eff. Size & \\ 
\midrule
Time & 0.93(0.70) & 0.55(0.53) & 3.11 & 0.002 & ** & \textbf{0.60} & Acc. \\
Location & 0.66(0.89) & 0.67(0.91) & 0.05 & 0.959 & - & 0.004 & Rej. \\
Emotion & 0.48(0.73) & 0.36(0.64) & 0.77 & 0.440 & - & 0.17 & Rej. \\
People & 0.83(0.75) & 0.80(0.79) & 0.30 & 0.766 & - & 0.04 & Rej. \\
Activity & 1.23(0.44) & 1.17(0.42) & 0.57 & 0.566 & - & 0.12 & Rej. \\
\midrule
Total & 4.13(1.83) & 3.55(1.74) & 1.83 & 0.067 & + & \textbf{0.32} & Rej. \\
\bottomrule
\end{tabular}
\end{table*}

\begin{table*}[ht!]
    \caption{
    The statistical analysis of the amount of contextual information contained in diary entries for the Baseline system between two counterbalanced groups using the Wilcoxon rank-sum test, where the p-value (-: p > .100, +: .050 < p < .100, *:p < .050, **:p < .010, ***:p < .001) is reported. Effect size with large or moderate magnitude is \textbf{bolded}.}
    \label{tab:hypo_test_info_amount_co_b}
    \centering  
\begin{tabular}{cccccccc} 
\toprule
\multirow{2}{*}{Category} & G1 Baseline & G2 Baseline & \multicolumn{4}{c}{Statistics} & \multirow{2}{*}{Hypothesis} \\ 
\cline{4-7}
 & Mean(S.D.) & Mean(S.D.) & W & p-value & Sig. & Eff. Size & \\ 
\midrule
Time & 0.97(0.71) & 0.86(0.69) & 0.64 & 0.525 & - & 0.16 & Rej. \\
Location & 0.75(0.92) & 0.46(0.78) & 1.18 & 0.240 & - & \textbf{0.32} & Rej. \\
Emotion & 0.48(0.79) & 0.46(0.57) & 0.46 & 0.638 & - & 0.03 & Rej. \\
People & 0.83(0.78) & 0.82(0.66) & 0.08 & 0.936 & - & 0.01 & Rej. \\
Activity & 1.33(0.47) & 1.00(0.27) & 2.41 & 0.016 & * & \textbf{0.78} & Acc. \\
\midrule
Total & 4.36(1.96) & 3.61(1.32) & 1.52 & 0.129 & - & \textbf{0.42} & Rej. \\
\bottomrule
\end{tabular}
\end{table*}

\begin{table*}[ht!]
    \caption{
    The statistical analysis of the amount of contextual information contained in diary entries for the \tool system between two counterbalanced groups using the Wilcoxon rank-sum test, where the p-value (-: p > .100, +: .050 < p < .100, *:p < .050, **:p < .010, ***:p < .001) is reported. Effect size with large or moderate magnitude is \textbf{bolded}.}
    \label{tab:hypo_test_info_amount_co_d}
    \centering  
\begin{tabular}{cccccccc} 
\toprule
\multirow{2}{*}{Category} & G1 \tool & G2 \tool & \multicolumn{4}{c}{Statistics} & \multirow{2}{*}{Hypothesis} \\ 
\cline{4-7}
 & Mean(S.D.) & Mean(S.D.) & W & p-value & Sig. & Eff. Size & \\ 
\midrule
Time & 0.63(0.48) & 0.47(0.55) & 1.21 & 0.228 & - & \textbf{0.30} & Rej. \\
Location & 0.97(0.97) & 0.35(0.72) & 2.33 & 0.020 & * & \textbf{0.71} & Acc. \\
Emotion & 0.31(0.67) & 0.41(0.60) & 0.89 & 0.374 & - & 0.15 & Rej. \\
People & 0.86(0.72) & 0.74(0.85) & 0.76 & 0.446 & - & 0.15 & Rej. \\
Activity & 1.11(0.32) & 1.24(0.49) & 0.89 & 0.374 & - & 0.29 & Rej. \\
\midrule
Total & 3.89(1.43) & 3.21(1.95) & 1.88 & 0.060 & + & \textbf{0.39} & Rej. \\
\bottomrule
\end{tabular}
\end{table*}

\begin{table*}[ht!]
    \caption{
    The statistical analysis of recall evaluation for the Baseline system between two counterbalanced groups using the Wilcoxon rank-sum test, where the p-value (-: p > .100, +: .050 < p < .100, *:p < .050, **:p < .010, ***:p < .001) is reported. Effect size with large or moderate magnitude is \textbf{bolded}.}
    \label{tab:co_recall_b}
    \centering  
\begin{tabular}{cccccccc} 
\toprule
\multirow{2}{*}{Category} & G1 Baseline & G2 Baseline & \multicolumn{4}{c}{Statistics} & \multirow{2}{*}{Hypothesis} \\ 
\cline{4-7}
 & Mean(S.D.) & Mean(S.D.) & W & p-value & Sig. & Eff. Size & \\ 
\midrule
Time & 1.07(0.69) & 1.00(0.57) & 0.52 & 0.603 & - & 0.11 & Rej. \\
Location & 0.93(0.86) & 0.59(0.72) & 1.73 & 0.084 & + & \textbf{0.40} & Rej. \\
Emotion & 0.76(0.86) & 0.73(0.76) & 0.03 & 0.973 & - & 0.04 & Rej. \\
People & 1.13(0.63) & 0.81(0.83) & 1.97 & 0.049 & * & \textbf{0.45} & Acc. \\
Activity & 1.62(0.49) & 1.49(0.50) & 1.11 & 0.268 & - & 0.26 & Rej. \\
\midrule
Total & 5.51(1.87) & 4.62(2.23) & 2.04 & 0.041 & * & \textbf{0.44} & Acc. \\
\bottomrule
\end{tabular}
\end{table*}

\begin{table*}[ht!]
    \caption{
    The statistical analysis of recall evaluation for the \tool system between two counterbalanced groups using the Wilcoxon rank-sum test, where the p-value (-: p > .100, +: .050 < p < .100, *:p < .050, **:p < .010, ***:p < .001) is reported. Effect size with large or moderate magnitude is \textbf{bolded}.}
    \label{tab:co_recall_d}
    \centering  
\begin{tabular}{cccccccc} 
\toprule
\multirow{2}{*}{Category} & G1 \tool & G2 \tool & \multicolumn{4}{c}{Statistics} & \multirow{2}{*}{Hypothesis} \\ 
\cline{4-7}
 & Mean(S.D.) & Mean(S.D.) & W & p-value & Sig. & Eff. Size & \\ 
\midrule
Time & 1.14(0.46) & 0.97(0.65) & 1.20 & 0.230 & - & 0.29 & Rej. \\
Location & 1.65(0.57) & 1.21(0.83) & 2.40 & 0.016 & * & \textbf{0.59} & Acc. \\
Emotion & 0.98(0.88) & 1.13(0.90) & 0.82 & 0.413 & - & 0.17 & Rej. \\
People & 1.51(0.50) & 1.33(0.76) & 0.79 & 0.428 & - & 0.27 & Rej. \\
Activity & 1.74(0.44) & 1.72(0.45) & 0.20 & 0.843 & - & 0.05 & Rej. \\
\midrule
Total & 7.02(1.36) & 6.36(1.87) & 1.69 & 0.091 & + & \textbf{0.39} & Rej. \\
\bottomrule
\end{tabular}
\end{table*}

\end{document}